# F and $V_k$ centres in $LaF_3$, $CeF_3$ crystals

E.Radzhabov and A.I.Nepomnyashikh

*Vinogradov Institute of Geochemistry, Academy of Sciences, Siberian Branch, Favorskii street 1a, P.O.Box 4019, 664033 Irkutsk, Russia, e-mail:eradzh@igc.irkutsk.su*

**Abstract.** Optical absorption properties of additively colored in calcium vapor and X-irradiated crystals with tysonite structure, $LaF_3$ and $CeF_3$, were studied in 1.5-5.5 eV energy range. Photobleaching with 1.9 eV light and dichroism of the absorption bands were investigated.

The absorption band at 3.75 eV with half width of 1.1 eV is observed in X-irradiated $LaF_3$, similar to the $V_k$-centre-bands observed in other fluorides. No such band is observed in X-irradiated $CeF_3$.

In additively colored $CeF_3$ absorption bands at 1.7 eV, 2.15 eV are observed. These bands can be attributed to F centres. Additive coloration upon heating in calcium vapor did not produce any new absorption band in $LaF_3$ crystals while formation of metallic inclusions, most likely La precipitates were observed. For X-irradiated $LaF_3$ and $CeF_3$, at temperature below 110K the photo-bleachable, dichroic bands are observed: near 2 eV -band perpendicular to c-axis, 2.6-2.7 eV -band along c-axis. F centres occupying the vacancies of fluoride F(1) layers in X-irradiated crystals and F-centres occupying all fluoride vacancies in additively colored crystals are proposed to be observed

**INTRODUCTION**

Lanthanide trifluorides have rather complicated crystal lattice of tysonite type. Each lanthanide ion surrounded by 11 fluorine ligands. There are three types of fluoride ions in the crystal cell, with different Madelung potentials (Van Gool *et al* 1969). F(1) fluorides build fluoride layers in the lattice; F(2), F(3) fluorides build mixed La-F layers. The lanthanide trifluorides containing Ce, Nd, Pr are the promising materials for scintillators. In spite of their chemical simplicity the defects created by ionizing radiation are not well understood.

F and $V_k$ centres in $LaF_3$, $CeF_3$ crystals were studied by ESR (Williams *et al* 1994). It was found 3-line hyperfine pattern in ESR of $LaF_3$ following X-ray irradiation, with g= 2.00 and total splitting of 1740 Gauss when B is along the c axis. These values correspond closely with the parameters of $V_k$ centres in other fluoride crystals and authors tentatively assign this as the $V_k$ resonance. This ESR signal anneals at 70 K (Williams *et al* 1995). ESR experiments on pure $CeF_3$ revealed clear spectra of the F centres in the X-irradiated crystals. There are two distinct local environments for the F centre yielding quartet and triplet hyperfine splitting, with different annealing temperatures of ~110 and ~140 K, possibly indicating impurity association (Williams *et al* 1994).

The optical absorption spectra of the $Co^{60}$ irradiated crystals at 77 K are formed by many absorption bands in the 1.5-4 eV range with broad maxima at 1.7 and 2.4 eV. The investigation of transformation of these centres shows that the observed absorption bands have to be assigned to electronic centres (Sheulin *et al* 1994).

In this paper the optical absorption spectra of F and $V_k$ centres as well as optical dichroism of their absorption were studied in $CeF_3$ and $LaF_3$. The photobleaching and optical dichroism experiment were successfully applied in the investigation of many color centers (Fowler *et.al.* 1968). They are very useful also for low symmetry crystals (for example F centres in BaFCl (Lefrant *et.al.* 1976)). Some preliminary results were published elsewhere (Radzhabov *et.al.* 1995).

**EXPERIMENTAL**

The crystals were grown by Stockbarger method. Purification from oxygen was obtained by addition of $PbF_2$ to the extent of about 1-3 wt. %. The mixture is slowly heated under vacuum up to melting point. Crystals are grown under vacuum $10^{-3} - 10^{-4}$ torr.

Both crystals have a rather low thermal conductivity. To obtain a low temperature of sample one needs a good thermal contact between them and sample holder. Particularly it is important under X-irradiation, when crystal surface is warm up by X-ray.

The crystals were additively colored in calcium vapor in vacuum above 700°C. The samples were placed into stainless steel tube and evacuated to near $10^{-2}$ torr. The 0.2 - 0.4 g of metallic calcium is placed into another boat. Then the samples were heated 2-3 hours at 700-950°C. Apart from alkaline-earth fluorides a metallic layer, probably calcium, forms on the surface of $LaF_3$ and $CeF_3$. Nevertheless both crystals were colored. The coloration depth was about 1-2 mm per hour at 750°C. After slowly cooling the samples were sawed and polished. Before measuring the additively colored crystals were annealed at 700 to 900°C for about 5 minutes and rapidly cooled in air. To avoid the contact with air under annealing the sample (particularly $CeF_3$) was surrounded by Cu foil. Nevertheless the additional polishing is necessary for $CeF_3$.

X-irradiation was produced by W tube operated at 40 kV, 40 mA. Absorption spectra were measured with spectrophotometer "Specord UV-Vis". Photobleaching was done at 80K by 200 W halogen lamp through filter with 1.9 eV high energy edge.

**RESULTS**

**Additive coloration**

Results of the coloration were sharply different in $LaF_3$ and $CeF_3$. $LaF_3$ crystals have a metallic shine after the coloration. Into microscope one can see a metallic inclusion more than 10 μm. These inclusions are probably precipitates of metallic La. No new absorption bands were seen in spectra of additively colored $LaF_3$. These metallic precipitates disappeared above 950°C but no absorption bands appeared after quenching. Apart from these we observed the appearance of F bands in additively colored $CaF_2$, $SrF_2$ crystals, which initially contained only F-centres aggregates.

$CeF_3$ crystals were blue colored after additive coloration. Apart from $LaF_3$ no precipitates were observed into microscope in these crystals. Two main absorption bands at 1.7 and 2.15 eV appeared in $CeF_3$ samples quenched from 900°C (Fig.1). After long storage at room temperature the bands at 2.5, 2.7 eV also are appeared. The additive coloration disappeared above 900°C.

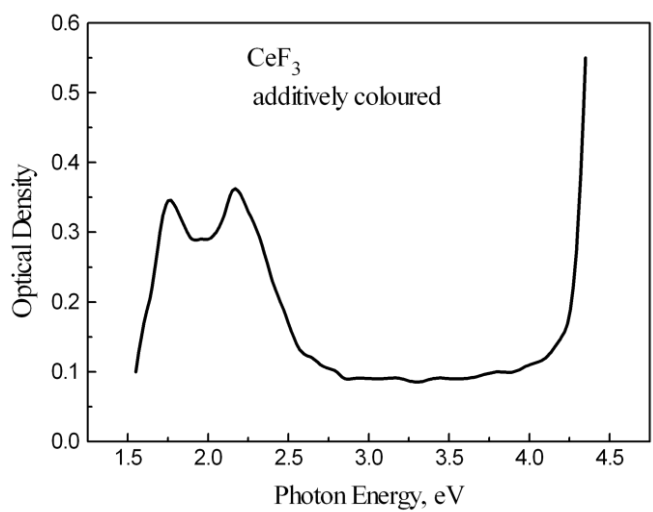

Fig 1. Optical absorption of additively colored $CeF_3$ crystal at room temperature.

**X-ray irradiation**

**LaF$_3$** The X-irradiation of the crystals leads to the creation of electronic centres as well as hole centres. Several absorption bands are created by X-irradiation at 80K (Fig.2). Near 100 K some of these bands disappear. The glow peak observed during this process.

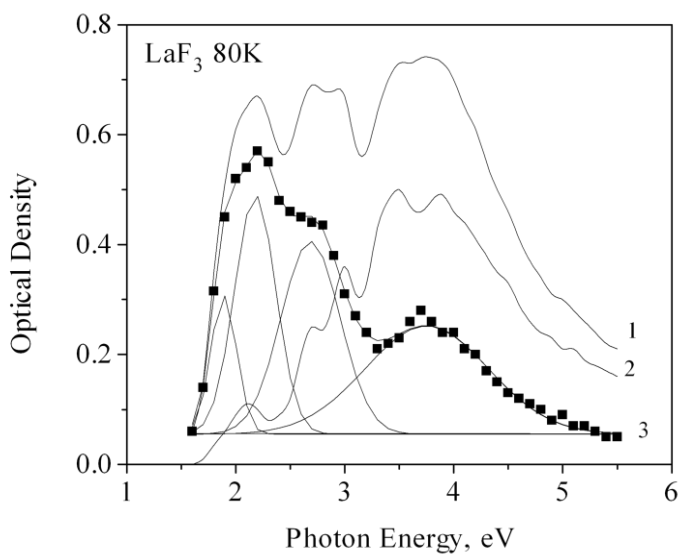

Fig. 2 Photobleaching of X-irradiated LaF$_3$ by 1.9 eV light at 80 K. Curve 1 - after X-irradiation , curve 2 - after subsequent 1.9 eV light irradiation, curve 3 is difference of curves 1 and 2. Curve 3 fitted by four gaussians (dashed curves).

The 3.75 eV band in LaF$_3$ was disappeared below 115 K. Moreover after X-ray irradiation the crystals had a slight blue luminescence in the dark. The 3.75 eV and other bands were slightly decreased with keeping at 80 K.

All other bands are extremely easy photobleached simultaneously by 1.9 eV light or light from 2.2 or 2.7 eV bands. Therefore all three bands belong to one type centre. The 1.9 and 2.1 eV centres are oriented perpendicular to the c-axis and 2.7 eV - along the c-axis (Fig.3).

**CeF$_3$** Some bands at 1.5-3 eV region disappear at 90-120 K temperatures in the X-irradiated CeF$_3$ crystals. The same bands can be easy photobleached by red light (Fig.4). Similar to LaF$_3$ the low energy transitions at 1.7, 2.2 eV are oriented perpendicular to the c- axis and 2.6 eV centres - along the c-axis (see Fig.4).

## DISCUSSION

**Additive coloration**
In many additively colored crystals the quenching procedure leads to appearance the single F centres rather than F centre aggregates. Therefore it may be suggested that the 1.7, 2.15 eV bands in quenched additively colored CeF$_3$ belong to F centres. It is well known that in additively colored alkali halides F-centres in thermal equilibrium are detectable only at sufficiently high temperatures (Zeitz 1954). At room temperature, the well-annealed crystals show only the colloidal bands of the alkali metal. F band in additively colored BaFCl crystal also was observed only above 320°C (Nicklaus *et al* 1972). We observe the F bands in additively colored CaF$_2$ and SrF$_2$ crystals above 200-300°C . Apart from this no absorption bands were observed in LaF$_3$ crystals up to 500-600°C . Evidently the thermal equilibrium of F-centres in alkali halides and alkaline-earth fluorides is described by the reaction:

$$F_n \Leftrightarrow F + F_{n-1} \qquad (1)$$

where n = 2,3,4 ... . The absence of the F absorption band means that the energy of F-centre thermal dissociation from colloids is higher than the energy of thermal ionization of F-centre. The metallic La precipitates begin to diminish in size above 900°C and dissolve above 950°C. Therefore the F-centre absorption cannot be obtained by this method in LaF$_3$.

There is obvious difference in the additive coloration of CeF$_3$ and LaF$_3$ crystals. At the similar thermal conditions the large metallic precipitates are formed in LaF$_3$, while the F centres and small F centres aggregate are formed in CeF$_3$ crystal. Possibly this difference is caused by different crystal structure of metallic La and Ce. The La metal has a hexagonal lattice, while the Ce metal has a face centered cubic structure (Kittel 1978). It seems likely that the formation of hexagonal La lattice from LaF$_3$ lattice needs a sufficiently less energy then the formation of Ce metal lattice in CeF$_3$. Let us consider the large aggregate of F centres in nearest fluoride sites. The result is equivalent to that when we remove all fluorides from part of lattice and add the subsequent amount of electrons (three electrons for one La). This leads to the formation of a hexagonal lattice of La atoms. This lattice is the same as crystal lattice of metallic La, while the distance between nearest atoms is somewhat larger (4.13 Å instead of 3.73 Å). We assume that the large metallic precipitates of La in additively colored LaF$_3$ is created because of similarity of La sublattice in LaF$_3$ and crystal lattice of La atoms. For the back process one needs to enlarge a La-La distance by melting the La precipitates. The metallic La is melted at 920°C. In line with this assumption we observe the decrease of La precipitate dimensions above 950°C.

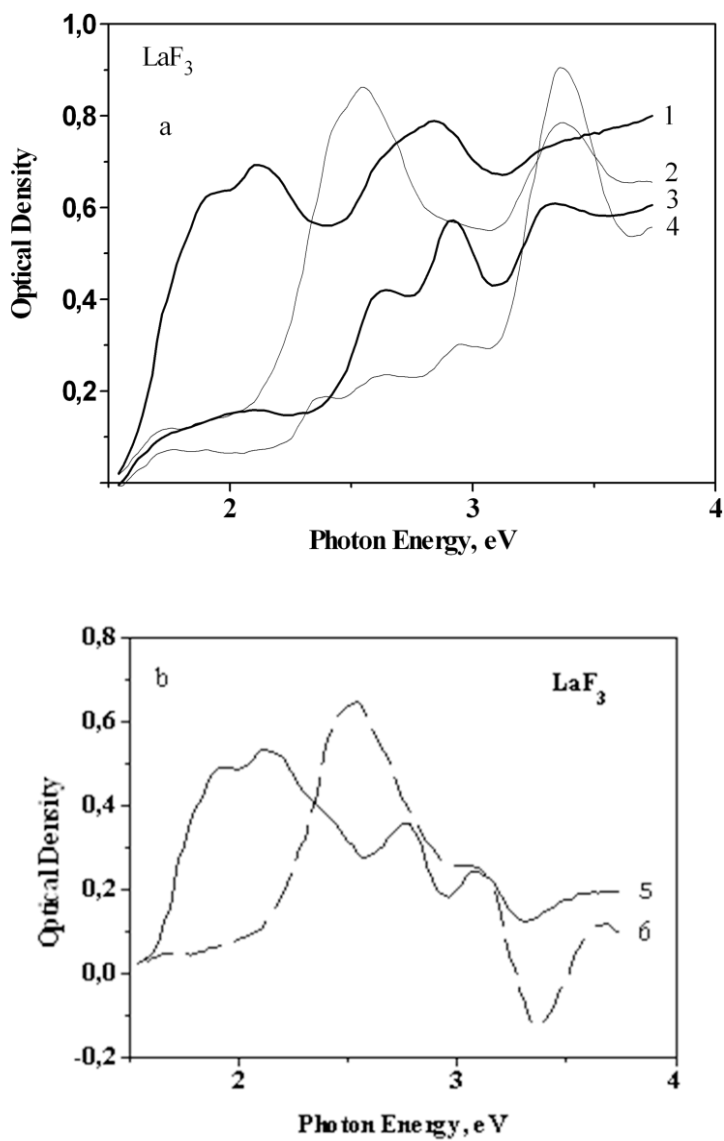

Fig 3. Dichroism of absorption of LaF$_3$ crystals at 80 K. Full line curves were measured perpendicular to c-axis, dashed or dotted curves - parallel to the c axis.
Plot a ; curve 1, 2 were measured after X-irradiation , curves 3 and 4 were measured after subsequent bleaching with 1.9 eV light. Plot b : curve 5 is difference of curves 1 and 3, curve 6 is difference of curves 2 and 4.

### X-ray irradiation

**LaF$_3$** The V$_k$ centre ESR spectrum in LaF$_3$ (Williams *et al* 1994) closely matched the spectra of V$_k$ centres in other fluoride crystals, except the unusually large halfwidth of V$_k$ lines. One can expect that the ultraviolet V$_k$ absorption band has to be in the range of 3.5-4 eV as in other fluorides. Therefore we assume that the absorption band at 3.75 eV with halfwidth 1.1 eV belongs to the V$_k$ centres.

The ESR spectrum attributed to V$_k$ centres in LaF$_3$ is annealed above 70K. No detectable change of absorption spectrum was observed at the V$_k$ centre annealing temperature of near 70K

(Williams et.al. 1995). Our experimental results show that the $V_k$ centres absorption band slowly annealed at 80K. Possibly at 80K experiments we see the $V_k$, which are stabilized by some defects.

We assume that the 1.9, 2.1 and 2.7 eV in X-irradiated $LaF_3$ are bands of the F centres as these bands are closed to F centres bands in $CeF_3$ and show similar dichroism.

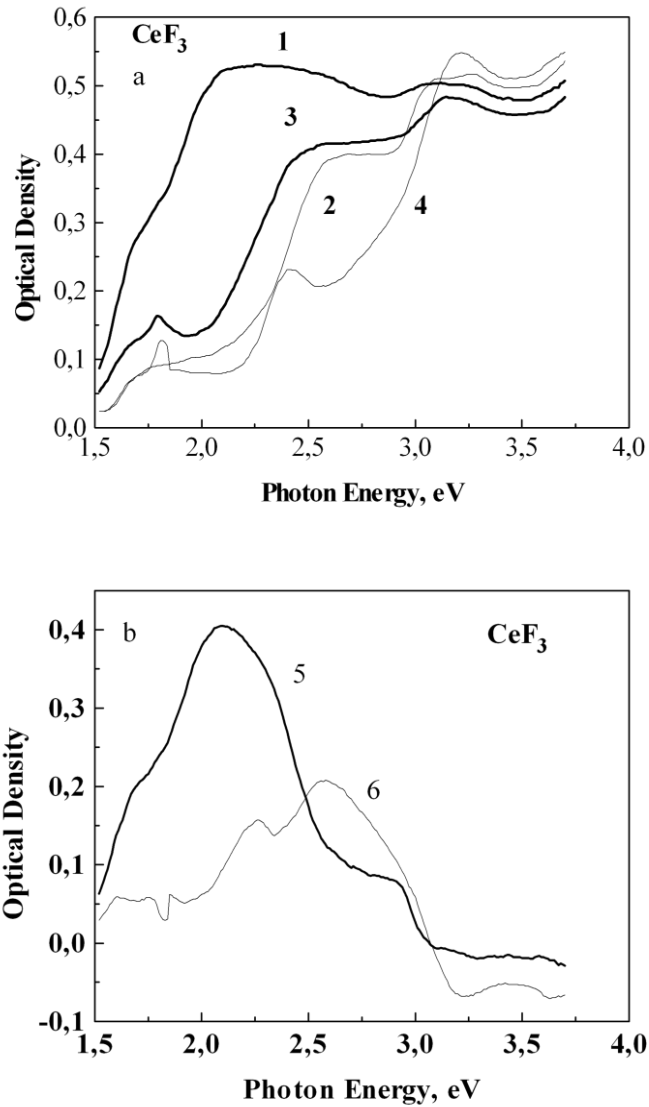

Fig. 4 Dichroism of F centres in X-irradiated $CeF_3$. Full line curves were measured perpendicular to c-axis, dashed or dotted curves - parallel to the c axis.
Plot a ; curve 1, 2 were measured after X-irradiation , curves 3 and 4 were measured after subsequent bleaching with 1.9 eV light.
Plot b : curve 5 is difference of curves 1 and 3, curve 6 is difference of curves 2 and 4.

$CeF_3$ No absorption band at $V_k$ region (3-4 eV) in $CeF_3$ is annealed within 80-120K. This is consistent with absence of the ESR $V_k$ spectrum of $CeF_3$ (Williams *et al* 1994).

The absence of $V_k$ centres in $CeF_3$ crystals is followed from our absorption spectra as well as from ESR spectra (Williams *et al* 1994). From X-ray photoemission spectra it is known that the 4f $Ce^{3+}$ ion levels lie in the band gap; about 6 eV lower conductivity band, while in the $CeF_4$ the 4f levels are empty (Kaindl *et al* 1987) as in $LaF_3$ crystals. The hole in $F^-$ 2p valence band is unstable and comes up to 4f level, creating the $Ce^{4+}$ ion. The oscillator strength for transitions from the $F^-$ 2p valence band to 4f cerium levels is likely to be very weak, because these transitions are not observed in absorption spectra of $LaF_3$. Therefore the hole from $F^-$ valence band is expected to rise slowly to 4f cerium levels. It is expected that the hole be self trapped before this transition. It seems the formation of transient self trapped holes can be observed in pulsed absorption or ESR spectra. The transient absorption spectrum in $CeF_3$ consists primary of the 2.1 eV peak which has attributed to F centres. It decays within 1 ms. No transient absorption at $V_k$ region was observed (Zhang et.al.1995). Obviously there is some fast process of hole transition from 2p fluoride zone to 4f cerium levels, possibly without self trapping.

The 1.7, 2.2 eV bands are rather close to 1.7, 2.15 eV bands of F centres in additively colored $CeF_3$ crystals. Additionally this bands annealed along with ESR of F centres in $CeF_3$ (Williams

*et al* 1994). The smaller 2.6 eV band which can be seen apparently only in dichroism spectrum (see Fig.4) also belong to this centres. Therefore the absorption bands at 1.7, 2.2 and smaller 2.6 eV band in X-irradiated $CeF_3$ belong to the F centres.

**F-centres absorption**

Although the actual unit cell has trigonal symmetry and six molecules, we shall make use of the simpler hexagonal unit cell with only two molecules (see also Olson *et al* 1978). Fluoride ions have three different symmetry positions in the lattice. F(1) fluorides build fluoride layers in the lattice; F(2), F(3) fluorides build mixed La-F layers. The F(2) and F(3) fluorides have the same positions in hexagonal unit cell.

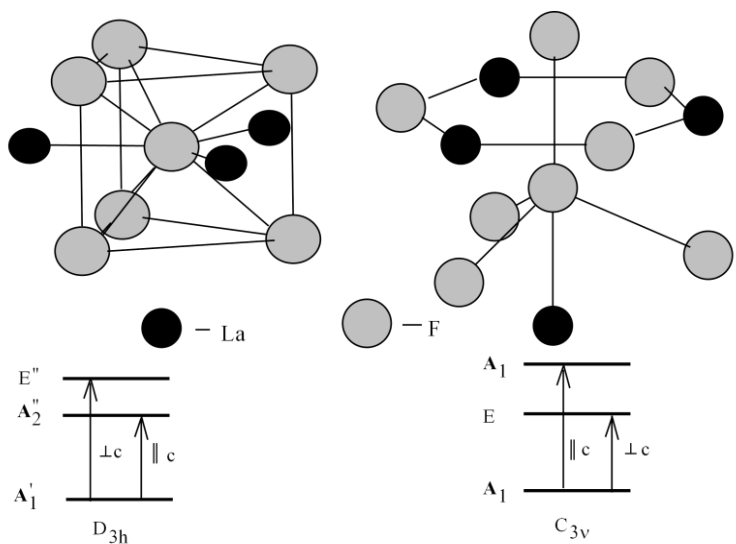

Fig. 5 Nearest environment for two fluoride sites of $LaF_3$ hexagonal lattice. The left picture belong to the F(2), F(3) fluoride sites from mixed layer. The right picture belong to F(1) sites from fluorine layer. The symmetry groups and possible order of energy levels are shown at the bottom.

The F(2) (or F(3)) and F(1) sites have a $D_{3h}$ and $C_{3v}$ symmetries respectively. Assume that z-axis orient along the c-axis. The irreducible representations for s, p atomic states of F centres in the fields of $D_{3h}$ and $C_{3v}$ symmetry are shown in the Table 1. Due to low symmetry of the crystal structure the 1s-2p absorption band of F centres has to be splitted into two bands. F centre has allowed $A_1 \rightarrow A_1, A_2''$ (E $\parallel$ c) and $A_1 \rightarrow E, E''$ (E $\perp$ c) transitions.

It is known from crystal field theory that overlapping of central ion orbital with negative ligands leads to increase of its energy (Murrell *et al* 1978). Owing to the calculation of F centres in BaFCl, SrFCl (Lefrant *et al* 1976) one can conclude that the overlapping with positive ion orbitals also leads to the similar result. Considering the $p_x$, $p_y$, $p_z$ orbital overlapping with nearest ions we can qualitatively estimate the order of A and E states for both fluoride sites. The $p_z$ orbital of F centre in mixed F(1), F(2) layers are not overlapped with surrounding ions (Fig.5). The $p_x$, $p_y$ orbitals sufficiently overlap with three $La^{3+}$ ions. Therefore the $E''$ level has to be at higher energy than $A_2''$ level. The reverse order of $A_1$ and E levels is estimated for F centre from fluoride F(1) layer. In this case the $p_z$ orbital are sufficiently overlapped with $La^{3+}$ (spaced at 2.35 A from F centre) and less overlapped with $F^-$ -ion (spaced at 2.64 A). The $p_x$, $p_y$ orbital have a small overlapping with three nearest fluorides (see Fig.5 right side). Therefore the $A_1$ level has to be at higher energy than E level. Experimental results are shown that in X-irradiated crystals F centers occupied one type of fluoride vacancies. From above rather sophisticated speculations one may conclude that the right picture ($C_{3v}$ symmetry) of Fig.5 is in line with observed spectra and we observed only F centres occupied the vacancies at fluoride F(1) layers.

The different symmetry of two fluoride sites in hexagonal $LaF_3$ lattice leads to the opposite dichroism of two absorption bands of F centres. The absorption bands of F centres in $LaF_3$ correspond closely with the F bands in $CeF_3$ (see Fig.3, Fig.4) although the F bands in $LaF_3$ are more resolved. The similar dichroism of the F centre bands in $LaF_3$ and $CeF_3$ indicates that F centres are located on one type of fluoride sites in both crystals. It seems the F(1) sites are preferred for F centres in LaF$_3$ lattice at 80 K. The Madelung electrostatic potential of F(2), F(3) sites in mixed layers of hexagonal lattice are higher by 1.2 eV than potential of F(1) fluorides (Van Gool *et al* 1969). The 2p fluoride valence band is expected to be split into two main groups due to this unequivalence (Olson *et al* 1978). One may assume that the highest Madelung potential of F(2), F(3) sites will be responsible for the localization of F centres. The Madelung potential of Br and F sites are 6.9 and 10.61 eV in BaFBr respectively (Baetzold 1992) and the F(F$^-$) centres (F centres on fluoride vacancies) in X-irradiated BaFCl are more thermally stable then the F(Cl$^-$) centres (Nicklaus *et al* 1972). Nevertheless in LaF$_3$ and CeF$_3$ crystals the F centres are created on fluorine sites with lesser Madelung potential. The possible explanation of this phenomenon is following. The diffusion energy of fluoride vacancy in F(1) layer is 0.2 eV, the energy for migration from F(1) to F(2), F(3) sites is 0.08 eV (Privalov et.al.1994). Usually the migration energy of the F centre is somewhat higher than that of the vacancies. Therefore we believe that at 80 K all F centres, which created at mixed layer sites, will diffuse into F(1) fluoride layer sites. To clear this problem one needs the further investigation as well as computer simulation of F centres in this lattice.

| Atomic states | $C_{3v}$ | $D_{3h}$ |
|---|---|---|
| s | $A_1$ | $A_1'$ |
| $p_z$ | $A_1$ | $A_2''$ |
| $p_x, p_y$ | E | E'' |

**Table 1**. The irreducible representations for s, p atomic states of F centres in the fields of $D_{3h}$ and $C_{3v}$ symmetry

There is difference between the apparent F center dichroism in X-irradiated LaF$_3$ and CeF$_3$ at low temperatures and absence of that in additively colored CeF$_3$ at room temperature. Obviously at low temperatures the F centres occupy only one type of fluoride sites while at high temperature the F centers occupy both fluoride sites.

| Crystal | Coloration procedure | Absorption, eV | Orientation to c-axis | T, K |
|---|---|---|---|---|
| CeF$_3$ | additive | 1.7, 2.15 | | 295 |
| CeF$_3$ | X-ray | 1.7, 2.2 2.6 | perpend. parallel | 80 |
| LaF$_3$ | X-ray | 1.9, 2.1 2.7 | perpend. parallel | 80 |

**Table 2.** Absorption bands of F centres in CeF$_3$ and LaF$_3$ crystals.

## CONCLUSION

The $V_k$ centres have 3.75 eV absorption band with halfwidth 1.1 eV at 80K. The centres were observed only in LaF$_3$.

F bands are shown in the Table 2. The F centres' transitions near 2 eV oriented perpendicular the c-axis and transitions near 2.7 eV oriented along the c-axis in X-irradiated LaF$_3$, CeF$_3$ crystals. It is assumed that the F centres occupy the vacancies at fluoride F(1) layers in X-irradiated crystals.

## ACKNOWLEDGMENTS

This work was partially supported by a program "Radiation Physics of Solid States" from State Committee of Education of Russia grant N 31222/L-4. The authors are grateful to Figura P., Mironenko S. and Krasin A. for preparation of crystals. They are also grateful to professor Williams R.T. for informing them of the results of a paper before publication.